\documentstyle[12pt,psfig]{article}
\begin{document}
\baselineskip=0.6cm
\newcommand{\EQ}{\begin{equation}}
\newcommand{\EN}{\end{equation}}
\newcommand{\EQA}{\begin{eqnarray}}
\newcommand{\EQN}{\end{eqnarray}}
\newcommand{\e}{{\rm e}}
\newcommand{\Sp}{{\rm Sp}}
\renewcommand{\theequation}{\arabic{section}.\arabic{equation}}
\newcommand{\Tr}{{\rm Tr}}
\renewcommand{\thesection}{\arabic{section}.}
\renewcommand{\thesubsection}{\arabic{section}.\arabic{subsection}}
\makeatletter
\def\section{\@startsection{section}{1}{\z@}{-3.5ex plus -1ex minus
 -.2ex}{2.3ex plus .2ex}{\large}}
\def\subsection{\@startsection{subsection}{2}{\z@}{-3.25ex plus -1ex minus
 -.2ex}{1.5ex plus .2ex}{\normalsize\it}}
\def\appendix{
\par
\setcounter{section}{0}
\setcounter{subsection}{0}
\def\thesection{\Alph{section}}}
\makeatother
\def\thefootnote{\fnsymbol{footnote}}

\makeatletter
\def\lesim{\mathrel{\mathpalette\gl@align<}}
\def\gtsim{\mathrel{\mathpalette\gl@align>}}
\def\gl@align#1#2{\lower.7ex\vbox{\baselineskip\z@skip\lineskip.2ex%
  \ialign{$\m@th#1\hfil##\hfil$\crcr#2\crcr\sim\crcr}}}
\makeatother
\begin{flushright}
hep-th/0004075\\
UT-KOMABA/00-08\\
April, 2000
\end{flushright}
\vspace{1cm}
\begin{center}
\Large
 STRING THEORY : Where are we now? 
 
\vspace{1cm}
\normalsize
{\sc Tamiaki Yoneya}
\footnote{
e-mail address:\ \ {\tt tam@hep1.c.u-tokyo.ac.jp}}
\\
\vspace{0.3cm}

{\it Institute of Physics, University of Tokyo\\
Komaba, Meguro-ku, 153 Tokyo}

\vspace{1cm}
Abstract
\end{center}
This is a brief overview on the current 
status of string theory for  non-specialists. The purpose is 
to give an aspect on the nature of string theory as a unified theory of all interactions 
including quantum 
gravity and to 
discuss future perspectives.  
Particular emphases are put on the mysteries why string 
theory contains gravity and why it resolves the ultraviolet 
problems. 

\section{History}
 
It has long been recognized that the two main  theoretical 
frameworks of present day physics,  quantum theory and relativity theory, are not easily reconcilable together 
microscopically.  Namely, the treatments of gravity using the methods of ordinary quantum field theory almost 
necessarily lead to the nonrenormalizable 
ultraviolet infinities.  
String theory is an attempt towards the ultimate 
theory which should explain all of the particle 
interactions and the fundamental structure of matter and 
space-time, 
by resolving the ultraviolet difficulties and the associated 
problems in a natural scheme where all other 
particle interactions can also be  taken into account 
in a completely unified manner. 
In this talk, I would like to explain why string theory 
is promising for this direction and what is the present 
status of the development of the theory. 

In view of the nature of this talk, it seems appropriate 
to start with some account of history.   Please refer to
 the table in the next page. 

The year 1998 was  the 30th anniversary of 
string theory \cite{review}.  The first clue for string theory came  
from the discovery \cite{veneziano} of simple formula for scattering amplitudes 
for hadrons (`strongly' interacting particles). They 
satisfy a duality symmetry, called the `$s$-$t$' duality 
at that time. This symmetry essentially says that 
we can represent the amplitudes symmetrically from 
the dual viewpoints of evolutions, {\it either} along the usual 
time-like direction {\it or} along space-like direction. 
Both give the same equivalent description 
of the physical scattering amplitudes. 
It soon turned out that such amplitudes are 
beautifully described by the quantum mechanics 
of relativistic strings \cite{nambu}.  In particular, the 
above $s$-$t$ duality is naturally encoded in the 
familiar mathematical 
properties of Riemann surfaces, which are interpreted as 
the base space for quantum mechanics of strings, 
namely, the two-dimensional 
field theory ({\sl conformal 
field theory}) describing the 
dynamics of the world sheet swept out by strings 
in the target space-time.  

After a few years of 
first explosion on the establishment of string theory 
as the theory of hadronic interactions,  
it was soon understood that the theory might 
rather be regarded as an extension of general relativity \cite{yo} \cite{ss} \cite{gso} 
and gauge theories (1972$\sim$ 1977) \cite{ns} \cite{yo2}.  However, it took a decade for 
particle physicists to recognize its significance 
as the guide for unified theory. The main reasons  
for this was that, at the same period, great developments  
are paralleled 
in renormalization theory of non-Abelian gauge theories 
which made us possible to describe the 
hadronic interactions in terms of ordinary local 
quantum field theories.  The successes prompted most 
particle physicists to further extensions of 
gauge field theories to unify all interactions including gravity. 
Such attempts culminated into the 
construction of theory of supergravity (1976$\sim$ 1980) \cite{peter} 
which generalized the general covariance 
to its supersymmetrically extended version.  
Actually, the most extreme of supergravity theory, $N=8$ supergravity in 
four dimensional space-time 
could also be understood as the dimensional 
reduction from 10 dimensional (super) string theory. 

\vspace{0.5cm}
\begin{center}
{\sl 
\begin{tabular}{l| c} 
1968 $\sim$ 1971 &   Veneziano, Virasoro-Shapiro formulae \\
&   String interpretation  \\
 &(Nambu-Goto action) \\
& (fermionic strings)\\
&\\
1973 $\sim$ 1976&  Relation with Yang-Mills theory \\ 
& and General Relativity \\
&  (Einstein theory and Supergravity) \\
& (ultraviolet finiteness)\\
& \\
1985 $\sim$ 1989 &   Classification of perturbative 
string \\
&theories in 10 space-time dimensions\\
& (I, IIA, IIB, H$_{SO(32)}$ and H$_{E_8\times E_8}$)\\
&\\
1994 $\sim$ present 
 &   Connection (`dualities') among \\
& perturbative string theories \\

&   `M-theory' conjecture \\
& (connection to 11D supergravity) \\
&   Role of `D-branes' \\
\end{tabular}
}
\vspace{1cm}
\end{center}

However, we soon understood that even supergravity 
could not resolve the ultraviolet difficulty of 
general relativity.  This is essentially due to the fact that 
supergravity is not unique if we include 
(space-time) higher derivatives. 
Namely, supersymmetry is not sufficient to control the 
short distance space-time structure and hence the 
ultraviolet divergences which are inherent to all 
local-field-theory approaches to gravity.  On the other hand, 
in the infrared, supersymmetry is very powerful. For instance, 
the classical action of N=8 supergravity is 
unique if we forbids higher derivatives than second in the 
field equation. 
 We  also understood that when one 
attempts to include the gauge interactions 
that could possibly fit in with the standard gauge models,  
one usually encounters various anomalies which 
violate the classical gauge symmetries and 
general covariance at the quantum level. 
It turned out \cite{gs} that the quantum anomalies of 
gauge symmetry and general covariance 
could be resolved in the field theories which could be 
regarded as the limit of supersymmetrical string-theory 
models.  

The last observation opened up the second explosion of string theories 
(1985 $\sim$ 1989), where the perturbative string theory 
models \cite{review} 
corresponding to stable perturbative vacua 
in flat 10
space-time dimensions are 
classified into five theories, type I $SO(32)$, type IIA, IIB, 
heterotic $SO(32)$ and $E_8\times E_8$. 
However, if one goes to lower dimensions by 
compactifying extra six space-time dimensions 
from 10 dimensions, it turns out that 
innumerable possibilities exist for the stable 
perturbative vacua. Thus the perturbative string theory 
has no predictive power for physics in 4
dimensions.  It is also noted that around from this period, 
the interests on string theory from the side of mathematicians  
arose. In particular, 
the conformal field theories with  a 
variety of nontrivial compactified 
target spaces provided 
various new interplays between mathematics 
and physics.  

During several years after the second explosion, 
some of the physicists have attempted to find the ways of 
formulating string theory in a nonperturbative fashion. 
For example, one such approach was to study certain toy models (called now `old' matrix models \cite{oldmatrix}),  which were 
soluble 
as string theory in lower dimensions, such as 
0+0, 1+0, 1+1 or even `negative' dimensional 
target space-times. It suggested some interesting 
hints on the structure of non-perturbative formulation, 
but unfortunately could not reach to 
spectacular successes from the original 
viewpoints of string unification. 
Around the same period, a different type of toy models 
became a focus of intensive studies,  namely the topological field theory.  Its physical significance is not clear.  On the mathematical side, however, the topological field theory provided new powerful 
methods in certain area of algebraic topology and/or geometry.  

Since around 1994 till present, we are in the third 
explosion of developments of string theory.  This began  
with the improved understandings \cite{dualityreview} 
on the 
relationship among the perturbative 
string vacua.  In particular, we are now gaining a 
good grasp on the relation among the 
five perturbatively consistent 
string theories. They are connected by various duality relations 
which exchange the regimes of weak and strong string couplings. 
 
\begin{figure}
\centerline{\psfig{figure=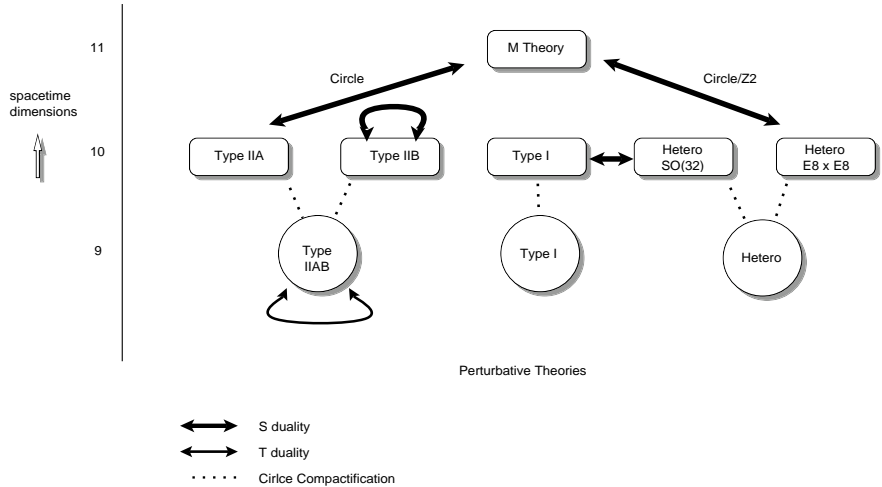,height=7cm,width=12cm,angle=0}}
\vglue-0.5cm
\caption{A schematic diagram showing various duality relations among perturbative string theories.  }
\label{rui}
\end{figure}

The  diagram above indicates the typical 
relationships.  
Here 
 nine dimensional theories in the third line 
of the diagram are 
obtained by the dimensional reduction 
of ten dimensional theories by compactifying one spatial 
dimension into a circle of radius $R$.  
The `S-duality' interchanges the strong and weak 
coupling regions $g_s \rightarrow 1/g_s$, while the `T-duality' reverses the 
radius of the compactification circle $R\rightarrow 1/R$ 
measured in the string unit where the fundamental 
string length parameter is set to one\footnote{
Do not confuse this S-and T- duality with the 
$s$-$t$ duality discussed before. 
}.  
A crucial assumption in this relation is the existence of 
`M-theory'\cite{witten}  such that it reduces to 10 
dimensional type IIA  (or Heterotic theory) 
by the dimensional reduction on a circle (or on a circle$/Z_2$) with radius $R\sim g_s \ell_s$. Also we found new degrees of freedom, 
Dirichlet-branes 
\cite{pol}, which are crucial building blocks to 
establish the above duality relation. They can 
be formulated as dynamical objects attached to 
the end points of open strings and behave as 
a sort of soliton-like excitations in string theory, 
which correspond to various monopole and instanton 
solutions in the low-energy field theory 
approximation to string theories.  

In view of this duality relation, we now believe that 
there must exist a unified theoretical framework 
in which all known perturbative string 
theories can be derived as `classical' 
solutions. 
In such a framework, we will perhaps be able to 
proceed to study the true dynamics of microscopic physics 
near the Planck length and thereby to give 
definite predictions.

 \section{Merits of string theory}
Now, is there any evidence for believing such a promise? 
Or is it merely a wild fancy of string physicists? 
Although it must be a long way to fix this question, 
we can at least mention the following points as merits 
or achievements of present string theory. 
\begin{enumerate}
\item Encompasses almost all past ideas towards 
unification of  particle interactions:

The past ideas include gauge invariance, Kaluza-Klein 
mechanism, supersymmetry, {\it etc}. 
\item Provides several new perspectives for understanding 
the dynamics of ordinary gauge field theories:

The most recent and remarkable example of this is the AdS/CFT correspondence \cite{adscft}, among others.

\item Provides  a realistically possible 
and conceptually satisfying scheme of unifying all interactions including gravity:

For example, the interaction and motion 
become a completely unified concept in string theory, and 
gravity is automatically included as an intrinsic 
property of its mathematical structure. 

\item Solves ultraviolet difficulty which is inherent to all the perturbative theories of particle theories with local interactions: 

Previous attempts to unify gravity suffer from the 
ultraviolet difficulty.  The removal of the ultraviolet 
difficulty within the usual framework of local 
field theory or in an extended framework allowing 
non-local interactions usually suffers from the violation 
of unitarity.  

\item Provides for the first time a microscopic 
explanation \cite{strovafa} of black hole entropy in terms of 
quantum statistical language: 

This is based upon the interpretation of the 
extremal and near extremal black holes 
in terms of Dirichlet-branes.  
The Dirichlet branes are the key for the most recent 
developments of string theory.

\end{enumerate}
    
The importance of resolving the ultraviolet 
problems with gravity being included can never be 
overemphasized. For example, 
if we try to compute the entropy in 
the usual local field theory, we necessarily 
encounter ultraviolet infinities, since the 
Newton constant is always infinitely renormalized. 
Not only that,  the renormalization also 
forces us to introduce infinitely many other 
dimensionful constants to write down the 
microscopic theory.   
 Remember that, in the history of quantum theory, 
the statistical interpretation 
of the entropy of black body played an indispensable 
role in identifying the correct 
microscopic degrees of freedom.  We have to remind ourselves that the ultraviolet catastrophe of classical field 
theory 100 years ago has never been completely 
resolved when we take into account gravity. 
Certainly, string theory provided the first (and only known)  promising direction toward its resolution. 

\section{Problems of string theory}
Although string theory is really promising in this way, 
 it is certainly true that the theory has its own problems 
in its present stage of developments.    String `theory' 
at present is merely a collection of rules of games for 
constructing scattering amplitudes ( elements of 
S-matrix)  using the 
various datum of conformal field theory. 
It is indeed an extension of the standard Feynman rules 
for constructing the scattering amplitudes 
in quantum field theory of particles based 
on perturbation theory. 
The datum for the particle Feynman rules are 
`propagators', describing the 
dynamics of world lines swept out by free particles 
in space-time, and `vertices' which describe 
the interaction, namely, the  transformations 
among particles in space-time  such as emission 
and absorption of particles.  In string theory, these datum are 
unified into conformal field theories on world sheets.  
The rules are astonishingly tight, 
self-consistent, and most importantly 
they conform to crucial 
physical requirements  for acceptable physical 
scattering amplitudes in general quantum theory, 
such as unitarity (conservation of probability). 
In particular, comparing with the 
particle Feynman rules, the string `Feynman rules' achieve  
a complete unification of free propagation and 
interactions of particles, while  in particle theory they 
must be given independently.  In other words, 
we would need, in general local field theory, separate principles 
for determining completely 
the particle spectrum and interactions. 
However,  we must admit that we have not yet 
arrived at a satisfactory understanding on  
why that works so well, why that 
conforms to general relativity at long distances, and 
what the basic principles are behind the rules. 
Worse than that, we cannot at present 
give definite physical predictions from string 
theory, because we do not know the real 
nonperturbative definition of string theory. 
Perhaps, our goal will be envisaged in the course for 
resolving these mysteries of present string theory.  

Therefore,  the most fundamental problem in 
string theory at present is to explore possible directions towards its nonperturbative formulation and the principles 
behind the rules based on which we can confidently 
construct the concrete mathematical framework.  

To explain the nature of such explorations, I will 
discuss some important mysteries, lying at the heart of 
physical 
properties of string theory rules,  
whose origin have not been understood completely 
even after the various surprising developments 
achieved in 30 years.  I hope that by so doing 
I might be able to 
convey some of the flavors to you on the matters 
what we are aiming to. 
 I will take two problems, first why string 
theory  contains  gravity, and secondly why string theory 
can resolve the ultraviolet infinities. 

\section{String to gravity}

Now, in what sense, does string theory contain gravity? 
Everyone here knows that gravity, as Einstein taught us, 
 is formulated as 
the space-time geometry based on (pseudo) Rimannian geometry of space-time.  In physical terms, this amounts to 
formulating gravity as a field theory of space-time metric and requires to treat all particle fields as geometrical 
objects (sections, connections, {\it etc}).  However, 
string theory in its present formulation does not 
require such geometrical objects at least at their starting point.  
Indeed, the usual formulation of string theory 
is done assuming just the flat space-time.  
Thus from the traditional point of view of dynamics, string theory, especially in its classical theory,  is merely describing the motion of  strings in flat 
space-time, and hence could not be the dynamical 
theory of space-time metric itself.  How can  string theory 
be the theory of gravity?

\subsection{A quantum physicists's derivation of general relativity}

To understand this, it is useful, before going directly to 
string theory, to take a brief digression 
on an elementary derivation of general relativity from a purely 
physical viewpoint of field theory without relying 
upon the Riemannian geometry.  

The basic idea of quantum field theory is that 
all the fundamental forces of nature can be 
understood as a result of exchanges of quanta 
corresponding to each force, such as photon 
(electromagnetic interaction), W-Z bosons 
(weak interactions), gluons (strong color force). 
The quantum of gravity is called graviton. 
The field theories of these quanta are constructed 
following the classic example of Maxwell theory. 
The photon is represented by the electromagnetic 
field or its vector potential $A_{\mu}(x)$ which universally 
couples to electric current $j_{\mu}(x)$,
\[
-\partial^2 A_{\mu}=e j_{\mu} , \quad \partial_{\mu}j_{\mu}=0 
\]
where $e$ is the unit of electric charge (or electrical 
coupling constant).  Here and in what follows, we 
will use the Euclidean conventions for the 
space-time indices. 
Unitarity requires that the only physical components 
of the vector potential are the transverse ones, 
since otherwise the time component of the 
vector potential yields negative probability 
according to usual probabilitistic interpretation 
of quantum theory. This leads 
to the gauge invariance requirement: Physical 
observables must be invariant under 
\EQ
A_{\mu} \rightarrow A_{\mu} + \partial_{\mu} \lambda 
\EN
with $\lambda=\lambda(x)$ is an arbitrary scalar. 
The field equation is then modified to 
\EQ
(-\partial^2 \delta_{\mu\nu} +\partial_{\mu}\partial_{\nu})
A_{\nu}=ej_{\mu} , \quad 
\EN
This is a consistent field equation as long as the current is conserved. 
The property of the electromagnetic force is 
precisely explained in this framework: For example, 
in the lowest order approximation with respect to 
$e$, the force is computed from the expectation value 
of the two-point correlator  $\langle (\int A_{\mu}j_{\mu})^2 \rangle$. 
Masslessness corresponds to the long-range nature 
of Coulomb force and nondefinite sign of the charge density 
$j_0$ to the existence of both repulsive and attractive 
forces. 

Now what is the corresponding construction for graviton? 
Since gravity is again a long-range force, the graviton  must 
corresponds to a field satisfying the same massless field 
equation on-shell. Also it is always attractive and 
couples universally to mass or energy-momentum. 
Only candidate for 
the currents leading to  
this property of universal gravitation 
 is the energy-momentum tensor $T_{\mu\nu}$, 
which is a second rank conserved tensor current.
\EQ
T_{\mu\nu}=T_{\nu\mu} ,\quad \partial_{\mu}T_{\mu\nu}=0 .
\label{energymomentumtensor}
\EN 
Thus the potential must also be a second-rank and symmetric 
tensor $h_{\mu\nu}$. 
\[
-\partial^2 h_{\mu\nu}=2\kappa^2 T_{\mu\nu} 
\]
where $\kappa^2$ is proportional to Newton's gravitational  constant. Unitarity again requires that only 
transverse components survive, which is ensured 
if the gauge invariance is assumed 
under 
\EQ
h_{\mu\nu}\rightarrow h_{\mu\nu} + \partial_{\mu}\lambda_{\nu} + \partial_{\nu}\lambda_{\mu}. 
\EN
This leads to the field equation
\EQ
-\partial^2 h_{\mu\lambda} -\partial_{\mu}\partial_{\lambda} h^{\nu}_{\nu} 
+\partial_{\nu}\partial_{\lambda}h_{\mu}^{\nu}
+\partial_{\mu}\partial_{\nu}h^{\nu}_{\lambda} -\eta_{\mu\lambda}(-\partial^2 h_{\nu}^{\nu} 
+ \partial_{\alpha}\partial_{\beta}h^{\alpha\beta}) 
=2\kappa^2 T_{\mu\nu} .
\label{hlinearfieldeq}
\EN  
This form is unique if we assume that the 
left hand side is of second order in the space-time 
derivatives and that the equation is Lorentz invariant. 
However, the result combined  with the conservation law 
(\ref{energymomentumtensor}) 
of the energy-momentum tensor is not completely consistent.  
The reason is that the graviton itself has nonzero energy. 
Hence, the exchange of the graviton induces the change 
of energy of matter and leads to the violation 
of the conservation law when we only take into account the matter energies.  But if we take into account the
energy and momentum of graviton to recover the conservation law, we have to include the graviton field to its 
second order in the energy-momentum tensor 
by adding the contribution of $h_{\mu\nu}$, $T_{\mu\nu} 
= T_{\mu\nu}^{matter} + T^h_{\mu\nu}$. 
This in turn modifies the equation of motion of the 
graviton itself and the 
gauge transformation too. The modification of 
graviton field equation leads to a further modification 
of the conservation law of the energy-momentum in 
the next order.  And hence we have  to again modify 
the energy-momentum tensor. This process continues to 
infinite order in the graviton field.  The field equation thus 
becomes non-polynomial with respect to the graviton field. 

It is an old common knowledge \cite{com} that the final result is 
nothing but the formal power series expansion obtained from 
 the 
Einstein field equation
$R_{\mu\nu} 
-{1\over 2}g_{\mu\nu} R = \kappa^2 T_{\mu\nu}$ 
by introducing the graviton field $h_{\mu\nu}$ as 
\EQ
g_{\mu\nu}=\delta_{\mu\nu} + \kappa h_{\mu\nu} 
\EN
 and the modified gauge transformation is 
equivalent to the transformation law of the 
metric $g_{\mu\nu}$ under the general coordinate transformation. 
To summarize, what we have seen is that, 
under a few reasonable assumptions, the requirement 
of getting consistent field theory for graviton 
could lead to general relativity even if we did not 
know the Riemannian geometry at all.  
It should be noted that the requirement of unitarity 
(or gauge invariance) 
and also the tacit assumption that the 
field equation contains the space-time derivatives 
only to second order are the essential assumptions 
for the above derivation.  

We have emphasized that there is a big 
conflict between quantum theory and 
general relativity, since it leads to 
ultraviolet difficulties. However,  it does not mean that 
both frameworks are {\it fundamentally} contradictory to 
each other.  In fact, at large distances, the situation is quite  contrary. For an example, 
an old story in the famous debates between Einstein and 
Bohr shows that the quantum mechanical 
uncertainty relation $\Delta E \Delta t \ge h$ 
is consistent with the equivalence principle when it 
applied to the measurement of weights  
in weak gravitational fields. 
Their mutual contradiction is manifested only at 
sufficiently short distances near the Planck length, 
where the quantum gravitational effects  
become of the same order as other nongravitational effects. 
The lesson we learn from this seemingly elementary 
discussion is that the geometric formulation 
of gravity at large distances could well be 
a natural  consequence of some well-defined microscopic 
theory of gravity which could possibly 
be based on new principles  entirely different from the 
ordinary Riemannian geometry, but 
 be perfectly consistent 
with the fundamental principles of both general relativity and quantum theory 
in the sufficiently large distance region.  This, I think, is 
a suggestive lesson in pursuing the unification of  
geometry and quantum theory.  From the viewpoint of 
quantum field theory, genuine observable quantities 
are only S-matrix elements.   Even the geometry itself 
must  ultimately be constructed from S-matrix,  
if we emphasize the operational aspect of any 
physical theory.  

\subsection{Why does string theory contain General Relativity ?}

 Let us now discuss why and how string theory 
contains general relativity. First we briefly summarize 
the string Feynman rules.  

\vspace{0.3cm}
\begin{center}
\underline{String Feynman Rules}
\end{center}

\begin{enumerate} 
\item string world sheet = Riemann surface :

\begin{center}
particle quantum mechanics 
\[\downarrow\]
 two-dimensional 
(super)conformal field theory
\end{center}

\item S-matrix : defined by the following path integral

\vspace{0.3cm}
$\Sigma$=\{Riemann surfaces\} 
$\rightarrow$ {\cal M} = (super) space-time 
\[\hspace{-1.5cm}
\sum_{\Sigma \rightarrow {\cal M}}\, \int_{{\cal M}}\,[dxd\psi]\, g_s^{-\chi(\Sigma)}
\exp\Bigl[ -{1\over 4\pi \alpha'}\int_{\Sigma}d^2\xi \, L(x,\partial_{\xi}x, \psi, \partial_{\xi}\psi,
\ldots)\Bigr] \prod_i  \int d^2 \xi_i \, V_i(\xi_i)
\]
\item world-sheet lagrangian\[
L = g_{\mu\nu}(x)\partial_{{\bar z}} x^{\mu}\partial_z x^{\nu} + \cdots
\]
\item (string length)$^2$ = 
$
\alpha' =\ell_s^2  
$

\item string coupling = 
$ 
g_s
$

\[
\chi(\Sigma) =2-2g -h-e
\]

$g$= \# of handles, \, \, $h$ =\# of holes , \, \, 
$e$=\# of external lines (punctures)

\end{enumerate}
\noindent
Here $x$ is the bosonic 
coordinate $x^{\mu}(\xi)$ and $\psi(\xi)$ collectively 
represents all the other world-sheet variables, and 
$z=\xi_1 + i \xi_2, \bar{z}=\xi_1 -i\xi_2$ are the 
homomorphic coordinates of a Riemann surface. 
 $g_{\mu\nu}$ is  the metric of the target space-time, 
which is  flat $g_{\mu\nu}=\delta_{\mu\nu}$ in 10 dimensional perturbation theory but 
is curved when we consider a compactified space-time.   The integral 
$\int_{{\cal M}}\,[dxd\psi]$ symbolizes the path-integral 
on a Riemann surface corresponding to a given 
topology of the world-sheet. The summation symbol 
$\sum_{\Sigma \rightarrow {\cal M}}$ means that 
the sum over all nonequivalent Riemann surfaces are made. 
The weight factor $g_s^{-\chi(\Sigma)}$ can actually 
be absorbed into the world-sheet Lagrangian by introducing the 
two-dimensional Einstein term ${1\over 4\pi}
\int d^2 \xi\, R^{(2)}\sqrt{g^{(2)}} \phi(x)$ coupled to 
the external dilaton field $\phi$ by the constant 
shift of the dilaton $\phi \rightarrow \phi + \log g_s$, 
where $g^{(2)}_{ab}$ is the intrinsic metric 
for two-dimensional world-sheet.  The summation over all 
topologies and over the moduli spaces of Riemann surfaces just 
fits to the requirement of unitarity 
of quantum theory. Namely, the singularities 
caused at the boundaries of the moduli space 
give the correct physical singularities of unitary amplitudes. 
The initial and final asymptotic states are represented by the 
product of 
vertex operators $V_i(\xi_i)$ which have one-to-one correspondence 
with the physical states which manifest themselves
 as the singularities 
at the boundaries of the moduli space of the Riemann surfaces. 

From the viewpoint of two-dimensional field theory, 
the rules are characterized by the local conformal 
invariance (or Weyl) invariance.  Namely, 
the theory is invariant under the Weyl transformation 
(and its supersymmetrical generalization) 
$g_{ab}^{(2)}(\xi)\rightarrow \rho(\xi) g_{ab}^{(2)}(\xi) $ 
of the intrinsic metric of the world sheet.  
Although the two-dimensional Einstein term apparently 
coupled with dilaton 
violates this symmetry, it actually is needed to cancel 
a quantum anomaly of the Weyl transformation 
associated with the vertex operator for dilaton, 
which is a massless scalar excitation contained 
inevitably in string theory and accompanied by the graviton. 
The requirement of local conformal invariance is the most 
crucial property of the string Feynman rules and leads to 
the following properties of the string S-matrix. 

\begin{enumerate}
\item Complete  unification of motion and interaction:

The particle spectrum and interactions are 
determined simultaneously, since locally there is no distinction between motion and interaction on the Riemann surface. 
\item Existence of massless spin 2 state = graviton : 

From the viewpoint of particle spectrum, this is schematically explained as follows. For simplicity, we explain it in the bosonic string theory. The case of theories with fermionic 
degrees of freedom is basically the same. 
The 
world sheet-conformal invariance 
leads to one constraint and one gauge freedom 
in each space-time dimensions for the bosonic 
coordinate for the left and right moving modes 
(namely, holomorphic and anti-holomorphic modes) 
separately. 
\[
x^{\mu}(\xi) = x^{\mu}(z) + \bar{x}^{\mu}({\bar z})
\]
\[
L(z)=(\partial_z x)^2+\cdots =0, \quad \bar{L}(\bar{z})=
(\partial_{\bar{z}}x)^2+\cdots=0
\]
\[
z\rightarrow f(z), \quad \bar{z} \rightarrow \bar{f}(\bar{z}) 
\]
This reduces the number of physical modes by two 
in each direction and hence the first orbital excitations  are 
decomposed into the irreducible components with 
respect to the rotation group O($D-2$) of transverse directions 
as 
\[
D^2 \rightarrow (D-2)^2 = {D(D-3)\over 2} 
\oplus {(D-2)(D-3)\over 2} \oplus  1 .
\]
In the language of relativistic quantum field theory 
they are represented by the massless 
symmetric tensor (graviton) $h_{\mu\nu}$, massless antisymmetric 
tensor $B_{\mu\nu}$ and massless scalar (dilaton) $\phi$ 
with gauge invariance under 
$\delta h_{\mu\nu}=\partial_{\mu}\lambda_{\nu} 
+ \partial \lambda_{\mu}$, 
$\delta B_{\mu\nu}=\partial_{\mu}B_{\nu} 
-\partial_{\nu}B_{\mu}$ and 
$\delta \phi =c $ with $c$ being a constant, respectively. 
Actually, because of the above mentioned anomaly, 
the gauge transformation of $\phi$ must be 
associated with the change of string coupling 
$\delta g_s =cg_s$.  

\item Background independence :

This comes about because there is a 
1  to 1 correspondence  between 
vertex operators and physical states of strings. 
In particular, 
deformation of background metric $g_{\mu\nu}$ is 
absorbed by the condensation of graviton modes
graviton vertex operator :  $h_{\mu\nu}(x)\partial_{\bar{z}}x^{\mu}
\partial_{z}x^{\nu}$ 
for $\partial^2 h_{\mu\nu}=0=\partial_{\mu}h_{\mu\nu}$. 

This has the correct gauge symmetry property on shell. 
Namely the gauge variation 
\[
\delta h_{\mu\nu}(x)\partial_{\bar{z}}x^{\mu}
\partial_{z}x^{\nu}=\partial_z\lambda_{\mu}\partial_{\bar{z}}x^{\mu} + 
\partial_{\bar{z}}\lambda_{\mu}\partial_{z}x^{\mu}
\]
is a total derivative, due to the world-sheet 
equation of motion $\partial_z\partial_{\bar{z}} x=0$. 
Similarly the gauge symmetry is also valid for the 
antisymmetric tensor $B_{\mu\nu}$ too.

More generally, all possible deformations of the world sheet lagrangian (including the boundary conditions 
in the case of open strings) are absorbed by the condensation of various modes of string theory (if  D-branes are taken into account). 

Thus string theory automatically is a quantum-dynamical 
theory of space-time, although it has not been 
constructed as such.  Consistency with unitarity and 
Lorentz invariance ensures that the theory is bound to be 
consistent with General Relativity at long-distance regime, 
provided that the low-energy limit is described by 
local field theory. The last point is connected with the next 
item. 
We have used the terminology `background' independence. 
This is {\it not} meant  that the present string theory is already formulated in a completely background independent manner.  But the structure of the theory locally  in an 
infinitesimal neighborhood of the theory space suggests that a truly background independent formulation 
should be possible.

\item The dynamics of strings is local with respect to world sheet: 

 In particular, the factorization 
property of the world sheet is satisfied. 
Namely, if we pinch off some cycles of the 
Riemann surfaces, the surfaces in general 
change topology or factorize into two disjoint 
surfaces. Since the dynamics of the 
world sheet is local, the dynamics also faithfully 
reflect the change of topology and disjointing of the surfaces. 
This implies that in the limit where the string length 
parameter $\ell_s =\sqrt{\alpha'}$ vanishes,  the string Feynman rules 
smoothly reduce to those of ordinary particle theories.  
Combining with some dimensional consideration, this leads to 
the conclusion  that the low-energy limit of the rules must be 
described by the local field theory 
with only two powers with respect to derivatives in 
any space-time dimensions.   
The explicit computations of scattering amplitudes 
involving gravitons have been carried out 
long time ago and provided confirmations of 
the general arguments \cite{yo} \cite{ss}.

\end{enumerate}

This is how string theory in general contains gravity 
(supergravity in supersymmetrical cases).  Most of you perhaps now understand why I have called the 
existence of gravity a {\it mystery} of string theory. 
Although we understood how gravity is contained 
at a `phenomenological' level, we do not have a 
`theoretical' explanation.  Why do the Weyl invariant 
string Feynman rules lead to gravity and 
other gauge forces at long-distances? There must be 
some fundamental mathematical formulation 
which will give `geometrical' explanations 
on this surprising phenomenon.   That would provide proper  principles based on which string theory 
is reformulated in a completely nonperturbative and 
background independent fashion.  

\section{Resolution of UV divergencies}

The next mystery I would like to discuss is why the ultraviolet 
difficulties are resolved \cite{shapiro} in string theory.  
This again is 
regarded as a consequence of the conformal invariance 
of the string Feynman rules.  Basically, there is 
no ultraviolet region in those rules, since 
the singularities of the moduli space of 
Riemann surfaces only reside on its 
boundaries. However, as we have already mentioned, 
the singularities of the scattering amplitudes 
caused at the boundaries of the moduli space are 
mostly physical ones required by unitarity.  
In contrast to this, the particle world-lines 
have extra singularities which do not 
corresponds to the boundary of the moduli space 
of Riemann surfaces.  Those are  points where the 
proper time of propagation of  particles vanishes. 
For any finite string length parameter $\ell_s$, those 
singularities are resolved in the Riemann surface 
because of the conformal invariance. A short time propagation 
with respect to some direction is actually a 
long-distance propagation of strings in another 
direction. 

Actually, there is a danger at the boundary of the moduli 
space that some additional singularities may 
arise which are not compatible with unitarity. 
This is the possible problem of tachyon divergencies. 
If the theory contains tachyons, they yield 
exponential divergences associated with infinitely long 
proper time.  That would be an indication of the 
instability of the vacuum one starts with in 
formulating the Feynman rules.  This is precisely where 
the supersymmetry in space-time plays a proper role. 
If the supersymmetry is realized linearly, 
there can be no tachyonic excitation. 
As emphasized already, the supersymmetry is 
not powerful enough to control the ultraviolet divergences. 
Its proper position in string theory is rather in 
the infrared region to ensure the stability of 
perturbative vacua. 

 Now we have understood that both the existence of 
gravity and the resolution of ultraviolet difficulties 
usually associated with gravity in the ordinary 
framework of local field theory rely upon 
the conformal invariance of the string 
Feynman rules.  This strongly suggests that 
there is some geometrical meaning for the
world-sheet conformal invariance. 
From the point of view of two-dimensional field theories 
in general, choosing the conformal invariant 
field theories amounts to considering only a 
very special class of 2d field theories.  
In terms of the language of 
renormalization group, we are considering the 
fixed-point theories.  This reminds us an analogy with 
the experience 
which physicists have had in early quantum theory in the beginning 
of the 20th century.  To explain the spectrum 
of hydrogen atom in terms of classical Newtonian 
mechanics, we have to choose only special 
orbits for an electron circulating 
around the nucleus  by imposing the Bohr-Sommerfeld  
condition. The latter condition is characterized by 
adiabatic invariance.  
It is tempting to compare this situation with 
the choice of fixed point theories characterized by 
conformal invariance in string theory. 
Then, just like there is the 
uncertainty relation behind the Bohr-Sommerfeld condition, 
it is natural to suppose the existence of some 
characteristic new relation by which the old 
theory is limited and the structure of new theory 
is signified.  
In this sense, the relation should contain the 
string length parameter $\alpha'$ which is 
the only free (but fundamental) parameter of string theory. 
Remember that the string coupling $g_s$ cannot 
be regarded as fundamental, since 
it can be absorbed by a constant shift of the dilaton field 
$\phi$. 

There is a very simple candidate relation which nicely fits to  the above expectations.  That is called the 
space-time uncertainty relation \cite{stu} \cite{stu2} 
\EQ
\triangle T \triangle X  \gtsim \ell_s^2
\EN
between an invariant characteristic length scale $\triangle T$
measured in the time-like direction 
and an invariant characteristic length $\triangle X$
measured in the space-like direction. This relation 
faithfully represents the manner by which the ultraviolet 
divergencies are eliminated in the string Feynman rules. 
Moreover, it turns out that the relation gives 
nice qualitative explanations of the properties 
of the interaction of D-branes.  For example, 
the relation leads to the prediction of the 
characteristic distance probed 
by the scattering of D-particles as 
given by $\triangle X \sim g_s^{1/3}\ell_s$ \cite{liyo} 
which is supposed to be the characteristic distance 
scale of `M-theory'.  Since I have a plan of publishing 
a separate article in which the space-time uncertainty 
relation is reviewed extensively and developed further, 
I do not describe further details on this relation here. 
For previous reviews, see \cite{liyo2}. 

\section{Future of string theory}

I hope that the foregoing discussions 
were of some help to understand the nature 
of string theory and its mysterious which must be 
reformulated more appropriately in the future 
for the construction of really nonperturbative 
unified theory.    Let me finally provide some personal 
viewpoints on the future development of string theory. 
It seems to me that further 
clarifications of the 
mysteries emphasized above are important in exploring 
the unified theory of geometry and quantum theory,  which 
must necessarily be the unified theory of matter 
and space-time. Let us recall them again using a slightly 
different way of looking. 

\begin{enumerate}

\item Formulation of the space-time uncertainty relation 

The space-time uncertainty relation is still only 
a qualitative characterization  without firm 
mathematical formulation.  For example, we do not know 
the way for precisely defining the 
quantities $\triangle T, \triangle X$.  
Its original derivation \cite{stu2} was based on the theory 
of general conformal invariants known as 
`extremal length', but that is clearly 
insufficient for our present purpose 
since the Riemann surface is 
the perturbative concept. We are aiming 
the nonperturbative definition of the theory and 
therefore must find some reinterpretation 
of conformal invariance which does not require 
perturbation theory.  The following analogy with the 
phase space of classical dynamics and its quantum 
version might 
be of some help in pursuing along this direction: 

\begin{center}
\hspace{0.3cm} classical phase space   \,   $\longleftrightarrow$    classical space-time

\hspace{0.7cm} $\updownarrow  \hspace{3.8cm} \updownarrow$

\hspace{0.7cm} canonical structure   \hspace{0.2cm}  $\longleftrightarrow$    `conformal structure' 

\hspace{0.7cm} $\downarrow  \hspace{3.8cm} 
\downarrow$

\hspace{0.5cm} space of quantum state  $\longleftrightarrow$  `quantum space-time' 
\end{center}
The space-time uncertainty relation can be interpreted as
 an analog of the 
ordinary Heisenberg uncertainty relation in 
the transition from the second line to the third 
line  of the above diagram. 

This analogy obviously suggests us to try an
 algebraic characterization 
of the space-time uncertainty relation by elevating 
the space-time coordinates into some operator algebra. 
For such an attempt,  I would like refer the reader to 
\cite{yo3} . 
 
\item A new way of looking at gravity in string theory 

The advent of D-branes has provided a new data for 
considering the question of why gravity is contained 
in string theory.  The low-energy or low-velocity behavior 
of D-branes and their interactions are described by the 
Yang-Mills type matrix models with maximal supersymmetries.  Based on this, an interesting conjecture for 
the possible formulation of M-theory  has been 
proposed in \cite{bfss} \cite{suss}.  If this conjecture is 
true, the mystery concerning the existence of gravity 
is further enhanced. We cannot find 
any remnant of gauge invariance associated 
with graviton in the Yang-Mills degrees of freedom 
only, since the graviton in this case only 
appears as the $t$ channel effect. In the $s$ channel, 
there is no graviton which directly corresponds to 
the one exchanged in the $t$ channel, at least apparently. 
On the other hand, in matrix-theory interpretation 
\cite{bfss} \cite{suss}, 
we have the Kaluza-Klein excitation of graviton in 
the $s$ channel.  Thus, if we interpret the theory 
as being defined in 11 dimensional space-time as 
M-theory, the phenomenon of reproducing gravity 
only using the Yang-Mills degrees of freedom 
is not inconsistent with general properties of string 
theory.  Indeed, explicit computations 
shows that we can simulate the graviton 
exchanges including its first non-linear effect by 
using supersymmetric Yang-Mills theories. 
There remains 
many puzzles to be clarified further in the general 
matrix model  approaches. 
For fuller discussions, I would like to invite the readers  
  to our original papers \cite{oyo}, or to a 
previous review \cite{yrev} and the references therein.

\item Deeper understanding on the correspondence 
between classical supergravity and string theory

If the general structure of string theory as we understand at 
present is basically justified for 
the ultimate unification, the new theory 
is characterized by one and only one new constant, 
string length $\ell_s$.  In other words, if we take the long 
distance limit 
$\ell_s \rightarrow 0$, we must be able to recover the 
standard quantum theory of gauge fields 
and classical general relativity. 
The former is characterized by the Planck constant $\hbar$, while the latter is by the Newton constant $G_N$. 
In particular, the Newton constant remains finite 
in the limit of going to classical physics $\hbar 
\rightarrow 0$.  Therefore it is natural to 
adopt the unit where $G_N =1$. In this unit, 
the dimension of the Planck constant itself is equal to the 
square of length. Thus we must have 
\[
\hbar = k \ell_s^2
\]
where $k$ is a numerical constant.  In string theory, 
the constant $k$ is in principle calculable in terms of the 
expectation value of dilaton $k=k(\phi)$. 
Thus in the presence of gravity, the transition from 
classical physics to quantum physics is equivalent to 
the introduction of string length. In other words, 
the quantization is necessarily the quantization of 
space-time. This line of arguments further 
strengthens the point of view emphasized above and 
suggests the importance of establishing some 
definite `correspondence principle' 
between classical physics and string theory.

\end{enumerate}

 \section{Acknowledgements}
 I would like to thank the organizers of the workshop 
for inviting me to this
 interesting  workshop. 
The present work  is supported in part 
by Grant-in-Aid for Scientific  Research (No. 09640337) 
and Grant-in-Aid for International Scientific Research 
(Joint Research, No. 10044061) from the Ministry of  Education, Science and Culture. 


\end{document}